\documentclass{article}

\newdimen\footheight
\usepackage{mysprocl}

\usepackage{graphicx,epsfig, psfrag}

\usepackage{amsmath}

\newcommand{\lfrac}[2]{\textstyle\frac{#1}{#2}\displaystyle}

\newcommand{\ntrl}[1]{\tilde{\chi}^0_#1}
\newcommand{\chrg}[1]{\tilde{\chi}^+_#1}

\newcommand{\snu}{\tilde{\nu}_e}

\def\mwrap#1{#1}
\def\dZneuL{\mwrap{\delta Z_{\tilde\chi^0}}}
\def\dZfneuLl(#1#2){\mwrap{\left[\delta Z_{\tilde\chi^0}\right]_{#1#2}}}
\def\dZfneuLlC(#1#2){\mwrap{\left[\delta Z_{\tilde\chi^0}\right]^\ast_{#1#2}}}
\def\dZfneuRl(#1#2){\mwrap{\left[\delta Z^{\rm R}_{\tilde\chi^0}\right]_{#1#2}}}
\def\dZchaL{\mwrap{\delta Z^{\rm L}_{\tilde\chi^+}}}
\def\dZchaR{\mwrap{\delta Z^{\rm R}_{\tilde\chi^+}}}
\def\dZfchaLl(#1#2){\mwrap{\left[\delta Z^{\rm L}_{\tilde\chi^+}\right]_{#1#2}}}
\def\dZfchaRl(#1#2){\mwrap{\left[\delta Z^{\rm R}_{\tilde\chi^+}\right]_{#1#2}}}

\def\dZSfl(#1#2){\mwrap{\left[\delta Z^{\tilde f}\right]_{#1#2}}}

\def\dZZAl(#1#2){\mwrap{\left[\delta Z_{\rm ZA}\right]_{#1#2}}}

\title{
COMPLETE ONE-LOOP CALCULATIONS\\
IN THE CHARGINO/NEUTRALINO SECTOR OF THE MSSM
}

\author{\vspace{-0.2cm}T.~FRITZSCHE}
\address{Max-Planck-Institut f\"ur Physik (Werner-Heisenberg-Institut)\\
F\"ohringer Ring 6, D 80805 Munich, Germany
}

\begin{document}

\setlength{\abovedisplayskip}{ 12pt plus 3pt minus 10pt}
\setlength{\belowdisplayskip}{ 12pt plus 3pt minus 10pt}

{\hfill{\small \bf MPP-2004-106}\\[-0.2cm]}

\maketitle\abstracts{
  The present status of the calculation of
  radiative corrections to chargino and neutralino pair
  production processes in the MSSM is summarized.
  The main focus will be on the use of the on-shell renormalization
  scheme for charginos and neutralinos in conjunction with
  DR-bar parameters, such as those of the SPA conventions.
  Associated soft and hard bremsstrahlung and an appropriate separation of
  QED-like parts in the full one-loop contributions
  will be addressed.
}

\vspace*{-0.7cm}
\section{Introduction}

Precision measurements of masses, cross sections, and decay rates of
neutralinos and charginos are expected to become feasible at an
$e^+e^-$ linear collider\cite{TESLA}.
Charginos $\chrg{i} (i=1,2)$ and neutralinos $\ntrl{j} (j=1,...,4)$ 
are the mass eigenstates of the charged and neutral higgsino- and
gaugino-fields. 

This sector is controlled by only three MSSM-specific parameters: 
the soft-breaking gaugino-mass parameters $M_1$ and $M_2$, and the 
parameter $\mu$ of the MSSM-superpotential.
Thus, the MSSM predicts strong correlations within the 
rich set of observables in the chargino/neutralino sector.
Quantum corrections at the 1-loop level to all of these physical
quantities and their model-inherent correlations
are known to be large 
\cite{Pierce:1993gj,Eberl:2001eu,Fritzsche:2002bi,Diaz:1997kv,Blank,Oeller}
and are therefore important for phenomenological studies.

In the following, the basis of loop calculations in the
chargino/neutralino sector of the  CP-conserving MSSM 
will be outlined, using on-shell renormalization conditions.
The presentation is based on~\cite{Fritzsche:2002bi}, 
for other approaches see~\cite{others2}. 

\vspace*{-0.1cm}
\section{On-shell Renormalization}

Renormalization constants are introduced for the chargino mass matrix $X$
and for the chargino fields $\chrg{i}$ $(i=1,2)$ by the transformations
\begin{eqnarray}
X & \to & X + \delta X \;\;;\;\;
\left\{
\begin{array}{rcl}
\omega_L\,\chrg{i} & \to & 
\left[{\mathbf 1} + \lfrac{1}{2}\dZchaL\right]_{ij}\,\omega_L\,\chrg{j}
\\
\omega_R\,\chrg{i}  & \to &  
\left[{\mathbf 1}+\lfrac{1}{2}{\dZchaR}^\ast\right]_{ij}\,\omega_R\,\chrg{j}
\end{array}
\right\}
\, .
\label{eqn:RvRChar2}
\end{eqnarray}
The matrix $\delta X$ consists of the counterterms of the 
parameters $M_2$, $\mu$, $M_W$, and $\tan\beta$, forming 
the mass matrix $X$.
The field-renormalization constants 

\goodbreak

\noindent
$\dZchaL$ and $\dZchaR$ 
are general complex 2$\times$2 matrices.
The neutralino fields $\ntrl{i}$ $(i=1,\ldots,4)$ and the
mass-matrix $Y$ are renormalized by the substitutions
\begin{eqnarray}
Y & \to & Y + \delta Y \;\;;\;\;
\left\{
\begin{array}{rcl}
\omega_L\,\ntrl{i} & \to & 
\left[{\mathbf 1}+\lfrac{1}{2}\dZneuL\right]_{ij}\,\omega_L\,\ntrl{j}
\\
\omega_R\,\ntrl{i} & \to & 
\left[{\mathbf 1}+\lfrac{1}{2}{\dZneuL}^\ast\right]_{ij}\,\omega_R\,\ntrl{j}
\end{array}
\right\}
\;\;.
\label{eqn:RvRNeutr2b}\label{eqn:RvRNeutr2}
\end{eqnarray}
The counterterm matrix $\delta Y$ contains
the counterterms for $M_1$, $M_Z$, 
and the electroweak mixing angle $\theta_w$.
$\dZneuL$ is a general complex 4$\times$4 matrix. 

Using the on-shell approach of \cite{Fritzsche:2002bi},
the pole masses of the two charginos 
and
of one neutralino, $m_{\ntrl{1}}$, are considered as input parameters,
and are used to specify both, the values of the parameters $\mu, M_1,
M_2$, and the respective counterterms.
De-mixing conditions for the renormalized two-point vertex functions
with on-shell external momenta of charginos and neutralinos 
fix the non-diagonal entries of the matrices $\dZchaL$, $\dZchaR$, and
$\dZneuL$;  
their diagonal entries are determined by normalizing the residues of the
propagators to be unity.  

The sfermion sector is renormalized using the procedure of
\cite{Hollik:2003jj}, for another approach see \cite{others1}.
Renormalization of $\tan\beta$, as the ratio  
of the VEVs of the two Higgs fields, is done in the 
$\overline{\rm DR}$-scheme~\cite{tanbetarenormalization}.
The formal description of parameter and field renormalization
in the Stan\-dard-Model-like part of the MSSM is taken over
from~\cite{Denner:kt}.

\vspace*{-0.3cm}
\section{\boldmath$\overline{\rm DR}$ Versus \boldmath$\rm OS$ Scheme}
\label{sec:DR2OS}
\vspace*{-0.1cm}

In the $\overline{\rm DR}$ scheme, divergent 1-loop
quantities are renormalized by adding counterterms 
that are proportional to the divergent parts,
$ \frac{2}{\epsilon}-\gamma+\log 4\pi \, , $
of the vertex functions, regularized using the dimensional-reduction
method. 
As a consequence, physical observables depend 
on the scale $\mu_{\overline{\rm DR}}$.
Input variables in the $\overline{\rm DR}$ scheme are
a natural choice for GUT-inspired parameter sets (e.g.~in SUGRA
scenarios).
On the other hand, in the ${\rm OS}$ scheme the renormalization
constants are fixed at physical scales; observables are thus scale
independent. The ${\rm OS}$ scheme is convenient for
calculations of cross sections and decay rates, because 
masses at Born level and in higher order agree (with few exceptions),
holding the correct phase-space kinematics already in tree-level
calculations.  

The translation between $\overline{\rm DR}$ and OS parameters
for the quantities $\mu$, $M_2$, $M_1$ of the chargino/neutralino sector
is performed in two steps:
\begin{enumerate}
\item Using  $\mu$, $M_2$, and $M_1$ in the $\overline{\rm DR}$ scheme as
  a starting point,
  the pole masses of three particles, e.g.\  
  both charginos and the
  lightest neutralino, are calculated at the one-loop level.
\item From those three physical masses the corresponding parameters in
  the ${\rm OS}$ scheme are deduced, using tree-level relations (which
  are left unaltered by construction in the ${\rm OS}$ scheme).
\end{enumerate}

\section{\boldmath{$e^+e^-$} Production Cross Sections Of Charginos And
  Neutralinos} 

\vspace*{-0.1cm}
\subsection{Born amplitudes and virtual corrections}
\vspace*{-0.1cm}

At lowest order, 
the amplitude $\mathcal{M}$ for chargino-pair production can be
described by $s$-channel photon and $Z$-boson exchange and by
$t$-channel exchange of a scalar neutrino $\snu$.
In the case of neutralino pair production, there is no
photon exchange at tree level, and $t$-channel exchange 
is mediated by the two selectrons $\tilde e^s$, $s=1,2$.
Diagrams containing Higgs lines are always negligible.

This set of Born diagrams has to be dressed by the corresponding
loop contributions containing the full particle spectrum of the MSSM.
The renormalization constants determined in section 2 
are complete to deliver all
counterterms required for propagators and vertices appearing in
the amplitudes.

\vspace*{-0.1cm}
\subsection{Real photons and ``QED corrections''}
\label{subsec:Bremsstrahlung}
\vspace*{-0.1cm}

Virtual photons 
attached to external charged particles
give rise to infrared (IR) divergences in the loop diagrams.
An IR-finite result is obtained by adding real-photon bremsstrahlung
integrated over the photon phase space.
The sum of the one-loop contribution to the cross section,
$\sigma^{\rm virt}$, and the bremsstrahlung cross section, 
$\sigma^{\rm brems}$, is IR-finite.

 For cancellation of the IR divergence, it is convenient to split
 $\sigma^{\rm brems}$ into a (IR-divergent) soft part 
 and a (IR-finite) hard part, both  depending on 
 a  soft-photon cutoff $\Delta E$ for the energy of the radiated photon,
 \begin{eqnarray}
 \sigma^{\rm brems}  =  
 \sigma^{\rm soft}(\Delta E) + \sigma^{\rm hard}(\Delta E)
 \;\;.
 \end{eqnarray}

Due to supersymmetry-relations,
there is no diagrammatic way to disentangle QED-like photonic 
virtual contributions from MSSM-specific parts.
One can, however, isolate the universal and leading
QED terms in $\sigma^{\rm virt} + \sigma^{\rm soft}$
resulting from photons collinear to the incoming $e^{\pm}$,
which contain the large logarithm
$L_e =\log \frac{s}{m_e^2}$.
The separation
\begin{eqnarray}
\sigma^{\rm virt} + \sigma^{\rm soft}   = 
\tilde\sigma + \sigma_{\rm remainder}
\;,\;
\tilde\sigma  =
\frac{\alpha}{\pi}\,
\left[
(L_e-1)
\log\frac{4 \Delta E^2}{s} +
\frac{3}{2}\,L_e
\right]  \cdot \sigma^{\rm Born}
\;,
\label{eqn:QED:1}
\end{eqnarray}
identifies a one-loop contribution $\sigma_{\rm remainder}$ 
that is IR finite and free of large universal QED terms.
$\sigma_{\rm remainder}$ contains 
the MSSM-specific radiative corrections,
whereas the subtracted part $\tilde\sigma$ in~(\ref{eqn:QED:1})
together with the hard brems\-strah\-lung part 
from initial-state radiation, $\sigma^{\rm hard}_{\rm ISR}$, can be
considered as a contribution of the type ``QED corrections'',
\begin{eqnarray}
\sigma_{\rm QED} & = &
\sigma^{\rm hard}_{\rm ISR} + \tilde\sigma \, .
\label{eqn:QED:3}
\end{eqnarray}
It is independent of the auxiliary cut $\Delta E$, and it
contains all large logarithms from virtual, soft, and hard photons.

\subsection{Results}
\vspace*{-0.15cm}

For illustration,
the following figure contains the integrated cross sections
for $\tilde\chi^+_1 \tilde\chi^-_1$ 
production in $e^+e^-$ annihilation for unpolarized beams.
Besides Born cross section and full
1-loop result, also $\sigma_{\rm remainder}$
and $\sigma_{\rm QED}$ are shown. 
The calculation has been performed with the aid of 
{\tt FeynArts} and  {\tt FormCalc}\cite{feynarts}.\\[-0.3cm]

\begin{center}
{\large
\psfrag{Process}{$e^+e^- \to \tilde\chi^+_1 \, \tilde\chi^-_1 \, (\gamma)$}
\psfrag{Spb}{$\sigma/{\rm pb}$}
\psfrag{EGeV}{$\sqrt{s}/{\rm GeV}$}
\psfrag{Born}{$\sigma^{\rm Born}$}
\psfrag{Remainder}{$\sigma_{\rm remainder}$}
\psfrag{QED}{$\sigma_{\rm QED}$}
\psfrag{Complete}{$\sigma^{\rm 1-loop}$}
\scalebox{0.55}[0.48]{
  \epsfbox{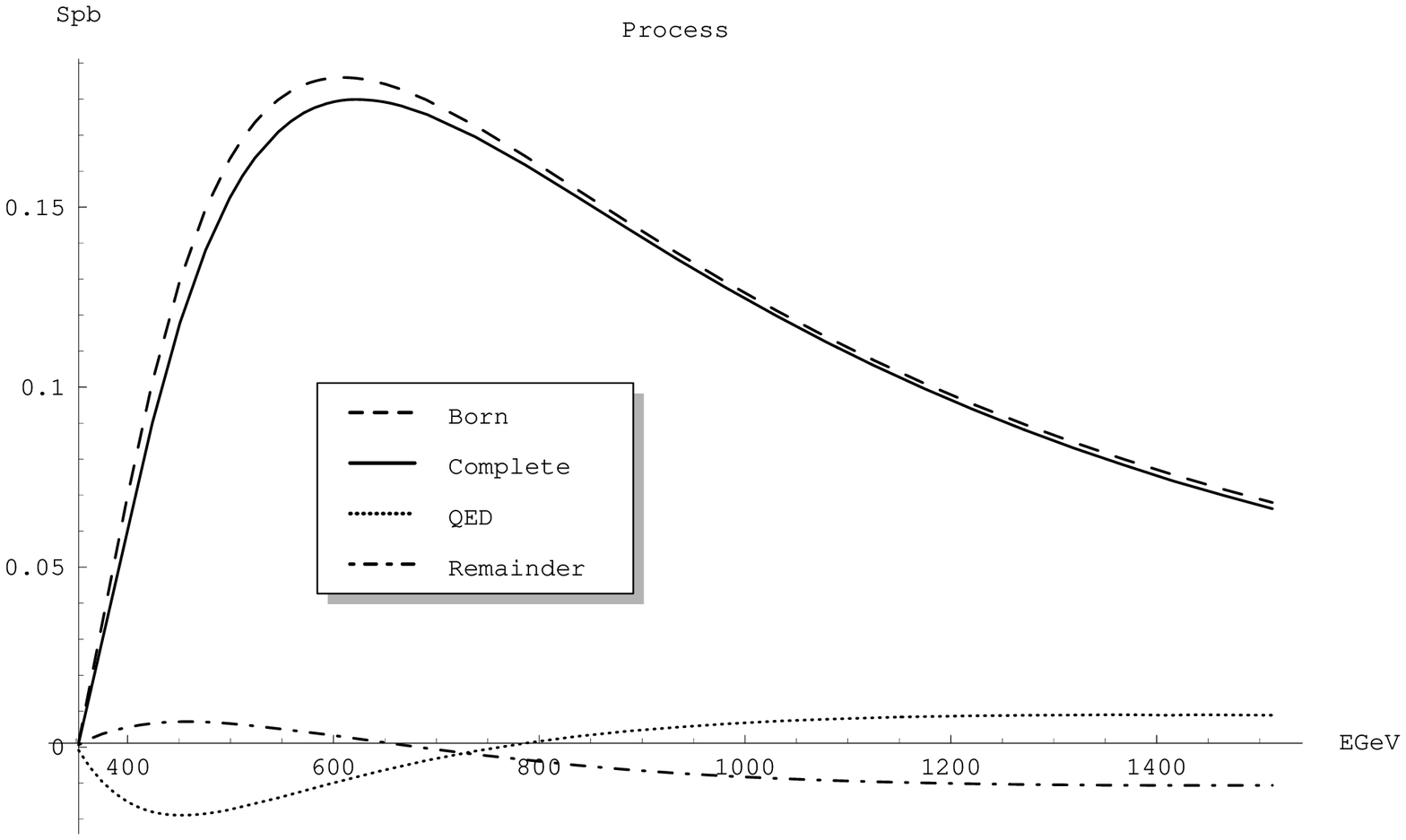}
}
}
\end{center}

\end{document}